\documentclass[prb,superscriptaddress,preprint,endfloats,showpacs]{revtex4}%preprint type

\newcommand {\IO}{IrO$_2$}
\newcommand {\TO}{TiO$_2$}

\newcommand {\Jhalf}{$J_{\mathrm{eff}}=1/2$}

\newcommand {\ttwog}{$t_{2\mathrm{g}}$}
\usepackage{graphicx}
\usepackage{bm} % bold math
\usepackage{pifont}
\usepackage{amsmath, amsthm, amssymb}
\usepackage{dcolumn} % Align table columns on decimal point
\usepackage{color}
\begin{document}
\title{Field-direction control of the type of charge carriers in nonsymmorphic {\IO}}
\author{M. Uchida}
\email[Author to whom correspondence should be addressed: ]{uchida@ap.t.u-tokyo.ac.jp}
\affiliation{Department of Applied Physics and Quantum-Phase Electronics Center (QPEC), University of Tokyo, Tokyo 113-8656, Japan}
\author{W. Sano}
\affiliation{Department of Applied Physics and Quantum-Phase Electronics Center (QPEC), University of Tokyo, Tokyo 113-8656, Japan}
\author{K. S. Takahashi}
\affiliation{RIKEN Center for Emergent Matter Science (CEMS), Wako 351-0198, Japan}
\author{T. Koretsune}
\affiliation{RIKEN Center for Emergent Matter Science (CEMS), Wako 351-0198, Japan}
\author{Y. Kozuka}
\affiliation{Department of Applied Physics and Quantum-Phase Electronics Center (QPEC), University of Tokyo, Tokyo 113-8656, Japan}
\author{R. Arita}
\affiliation{RIKEN Center for Emergent Matter Science (CEMS), Wako 351-0198, Japan}
\author{Y. Tokura}
\affiliation{Department of Applied Physics and Quantum-Phase Electronics Center (QPEC), University of Tokyo, Tokyo 113-8656, Japan}
\affiliation{RIKEN Center for Emergent Matter Science (CEMS), Wako 351-0198, Japan}
\author{M. Kawasaki}
\affiliation{Department of Applied Physics and Quantum-Phase Electronics Center (QPEC), University of Tokyo, Tokyo 113-8656, Japan}
\affiliation{RIKEN Center for Emergent Matter Science (CEMS), Wako 351-0198, Japan}

\begin{abstract}
In the quest for switching of the charge carrier type in conductive materials, we focus on nonsymmorphic crystals, which are expected to have highly anisotropic folded Fermi surfaces due to the symmetry requirements. Following simple tight-binding model simulation, we prepare nonsymmorphic {\IO} single-crystalline films with various growth orientations by molecular beam epitaxy, and systematically quantify their Hall effect for the corresponding field directions. The results clearly demonstrate that the dominant carrier type can be intrinsically controlled by the magnetic field direction, as also evidenced by first-principles calculations revealing nontrivial momentum dependence of the group velocity and mass tensor on the folded Fermi surfaces and its anisotropic nature for the field direction.

\end{abstract}
\maketitle

Band structure and its filling are fundamental factors determining electronic features of materials. The energy band is directly affected by some interactions such as electron correlation and spin-orbit coupling, which can give rise to distinct phases including Mott insulator and topological insulator depending on the filling. \cite{mott, ti1, ti2} Here we consider the electronic structures from a standpoint of the space group, especially unique dispersions and derived functions of nonsymmorphic materials, which possess symmetry operations of screw rotation and/or glide mirror. The nonsymmorphic crystals have been theoretically reexamined in recent years for possible nontrivial phases relevant to the nonsymmorphic crystalline symmetry. \cite{nonsymmorphic1, nonsymmorphic2} In nonsymmorphic lattices, for example, the ground state can be metal even at certain integer fillings, unless additional transitions are induced such as by the electron correlation or magnetic ordering. \cite{nonsymmorphic1} The nonsymmorphic symmetry operations stick bands together and provide additional degeneracies at high symmetry points, \cite{nonsymmorphictext} resulting in highly anisotropic folded Fermi surfaces near the Brillouin zone boundary. Therefore, the dominant type of charge carriers (electron or hole) is expected to be strongly dependent on the crystal orientation or capable of being switched by external stimuli. While the electric field control of the carrier type is usual in ambipolar semiconductors, \cite{ambipolar} control by other means remains largely unexplored.

\begin{figure}
\begin{center}
\includegraphics*[width=12.5cm]{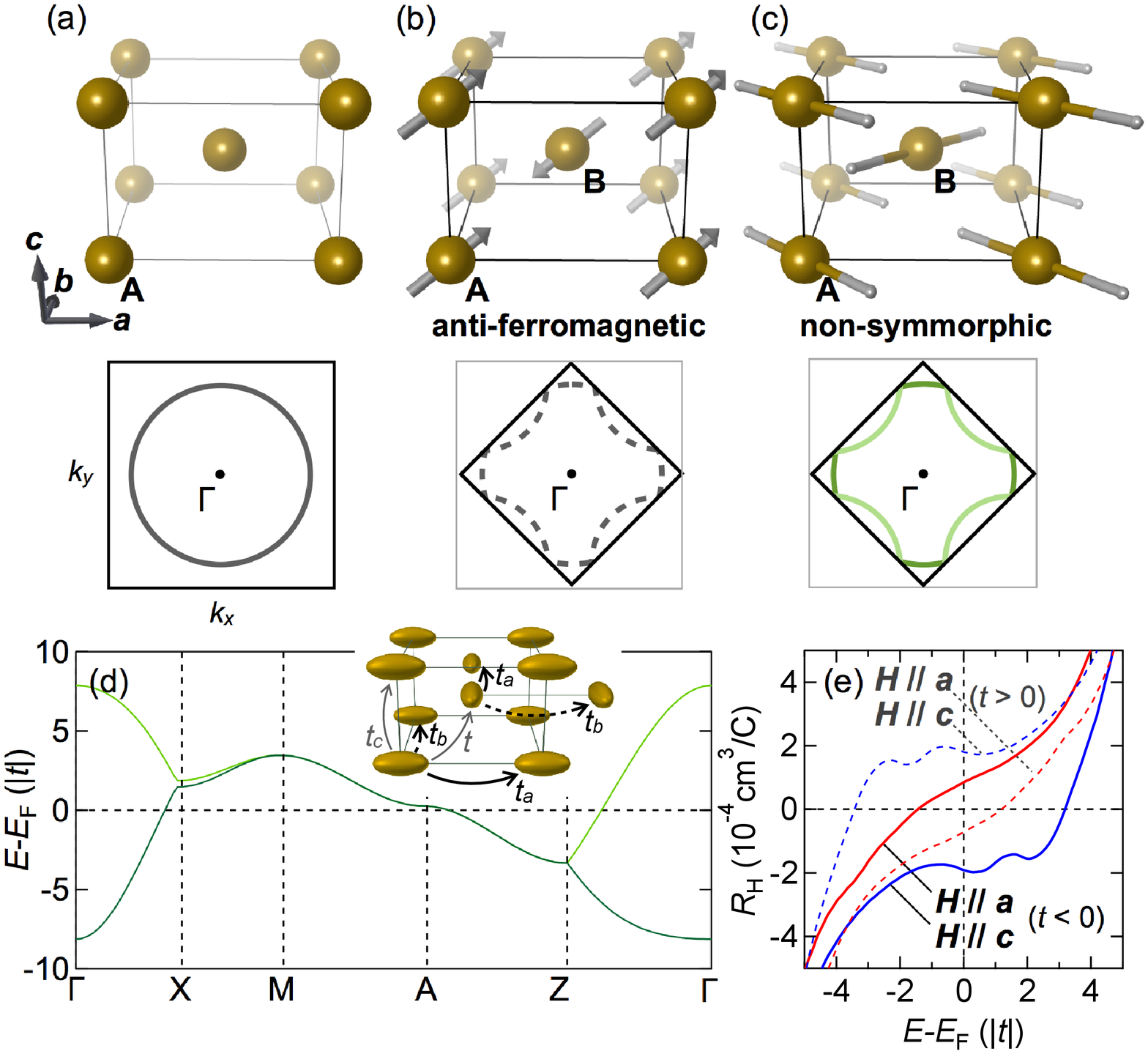}
\caption{
(Color online)
(a) Schematic Fermi surface at half-filling for the upper body-centered tetragonal lattice. (b) Antiferromagnetic spin configuration on sites A and B, and resultant gapped Fermi surface in the folded Brillouin zone. (c) Rutile structure as an example of nonsymmorphic lattices, where crystallographically nonequivalent sites A and B with opposite ligand configurations are interconnected by symmetry operations of screw rotation and glide mirror. Folded Fermi surfaces with strong anisotropy are then realized near the zone boundary without gap opening. (d) Energy bands calculated for a nonsymmorphic body-centered tetragonal model as shown in the inset, where an anisotropic orbital and its 90-degree rotated one are respectively placed on the two sites for reproducing the nonsymmorphicity. Tight-binding parameters $t_{a}=0.5t, t_{b}=0.4t,$ and $t_{c}=-0.8t$ ($t<0$) are used, and the Fermi energy $E_{\mathrm{F}}$ is defined as the energy of the half-filled level. (e) Hall coefficient depending on the field direction, derived in the tight-binding model for $t<0$ or $t>0$.
}
\label{fig1}
\end{center}
\end{figure}

Let us describe our idea of carrier switching in nonsymmorphic materials in more detail. Schematic in-plane Fermi surface at half-filling is first shown in Fig. 1(a), taking the example of body-centered tetragonal lattice. It is obviously difficult to control the carrier type or band topology of the large Fermi surface centered around the $\Gamma$ point. As shown in Fig. 1(b), Brillouin zone folding, e.g. associated with antiferromagnetic spin ordering on the two sites A and B, makes the small Fermi surface located around the zone boundary, but its electronic structure is usually gapped and the charge carrier is not defined. Figure 1(c) illustrates the rutile structure as an example of nonsymmorphic crystal, where sites A and B are crystallographically nonequivalent and related by the symmetry operations of screw rotation and/or glide mirror. Owing to the resultant zone folding, highly anisotropic and partly degenerate Fermi surfaces appear around the zone boundary without the gap opening.

Figure 1(d) exemplifies energy bands calculated with a simple but nonsymmorphic tight-binding model on the body-centered tetragonal lattice, where an anisotropic orbital and its 90-degree rotated one are respectively placed on the two sites to meet the nonsymmorphic symmetry requirements. The two semi-metallic bands stick together along M($\pi, \pi, 0$)--A($\pi, \pi, \pi$)--Z($0, 0, \pi$) lines on the zone boundary, due to the [001] fourfold screw axes (M--A) and the (110) glide planes (A--Z). \cite{nonsymmorphictext} As shown in Fig. 1(e), the Hall coefficient calculated with the tight-binding model is highly dependent on the magnetic field direction. Its sign is opposite between $H \parallel a$ and $H \parallel c$ around the half-filled level, suggesting that the carrier type probed by Hall effect can be surely controlled by the magnetic field direction. 

Here we adopt {\IO} for realizing the magnetic field switching of the carrier type. {\IO} is known to have the rutile structure, one of the most typical and simple nonsymmorphic crystals. In addition, {\IO} is a half-filled metal without showing any magnetic ordering nor other phase transitions down to low temperature, \cite{IrO2transport1} which is also critically important for our purpose. In recent years, iridium oxides have attracted increased attention due to the novel quantum phases originating from the strong spin-orbit interaction. \cite{SOI} The metallic ground state of {\IO} has also been intensively studied by advanced spectroscopies \cite{IrO2bandopt, IrO2calc2DFT, IrO2XPS1, IrO2XPS2, IrO2XAS, IrO2RXS} as well as first-principles calculations, \cite{IrO2calc1DFT, IrO2calc2DFT, IrO2calc3DMFT} while its theoretical pictures have been discussed through the ages. \cite{rutileoxides1, rutileoxides2, rutileoxides3} The low-energy electronic state can be represented by the half-filled state mainly with effective total angular momentum {\Jhalf}, \cite{IrO2XAS, IrO2RXS} in common with many insulating iridium complex oxides, though the band character itself is not essential for the carrier switching. Owing to this superior electric conductivity, polycrystalline and amorphous {\IO} films have been studied towards application to highly conductive electrode \cite{IrO2pld1, IrO2pld2, IrO2pld3, IrO2filmsputter1, IrO2filmsputter2, IrO2ald1, IrO2ald2} and efficient spin-current detector. \cite{IrO2SH} Here we prepare {\IO} single-crystalline films with various growth orientations by molecular beam epitaxy, and verify the predicted carrier control in nonsymmorphic materials by systematically measuring their Hall effect and comparing to the first-principles calculation results.

\begin{figure}
\begin{center}
\includegraphics*[width=12.5cm]{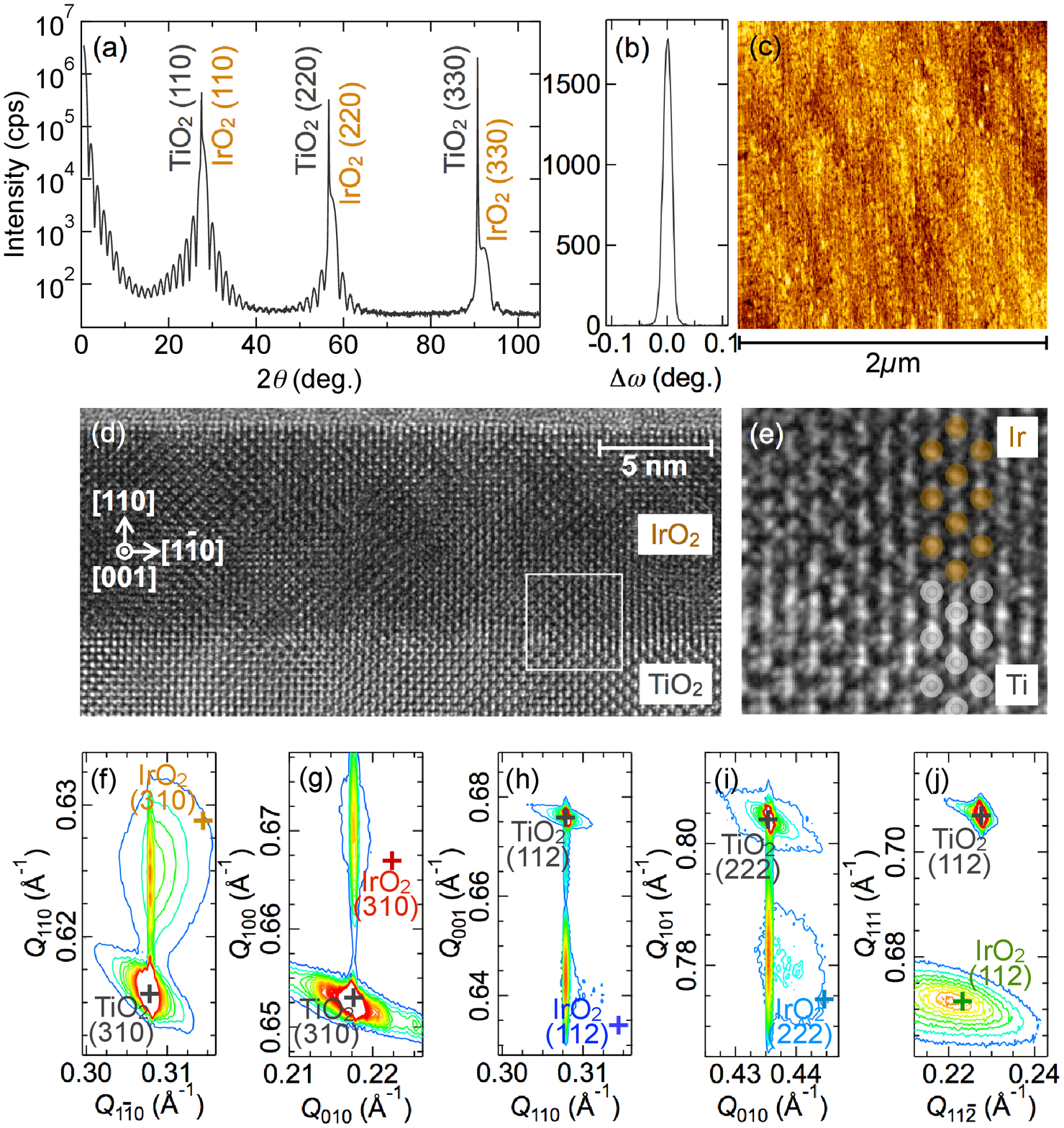}
\caption{
(Color online)
(a) X-ray diffraction $\theta$--2$\theta$ scan of an {\IO} film grown on (110) {\TO} substrate, and (b) rocking curve of the (110) film peak with a full width at half maximum (FWHM) of 65 arcseconds. (c) Atomic force microscopy image of the 10 nm (110) {\IO} film. The root mean square (RMS) roughness is 0.09 nm. (d) Cross-sectional high-resolution transmission electron microscopy image and (e) its magnified view around the interface. Reciprocal space mappings of (f) (110) and other (g) (100), (h) (001), (i) (101), (j) (111) orientated {\IO} films, epitaxially grown on the corresponding {\TO} orientations. Crosses denote the respective peak positions calculated from bulk lattice parameters. \cite{TiO2bulkstructure, IrO2bulkstructure}
}
\label{fig2}
\end{center}
\end{figure}

{\IO} films with growth orientations of (110), (100), (001), (101), and (111) were deposited on the corresponding {\TO} single crystal substrates using a Veeco GEN10 oxide molecular beam epitaxy system. Ir flux was supplied by electron beam evaporator. The deposition was performed in distilled pure ozone at a pressure of $1\times 10^{-6}$ Torr and at a substrate temperature of 300 $^{\circ}$C. Under these conditions, stoichiometry difference among the films should be negligibly small. \cite{supplemental} Film characterization is exemplified in Figs. 2(a)-2(e) for a (110) film, conveying the high crystallinity and excellent flatness. Single-phase film peaks of {\IO} are observed in the x-ray diffraction scans, with clear Kiessig fringes and sharp rocking curves (Figs. 2(a) and 2(b)). A step-and-terrace-like surface morphology is confirmed in the atomic force microscopy topography (Fig. 2(c)). Crystal defects are not substantially detected in the atomically resolved transmission electron microscopy images (Figs. 2(d) and 2(e)). Figures 2(f)-2(i) summarize the reciprocal space mappings of the complete set of samples fabricated, demonstrating their high-quality epitaxial growth on the corresponding {\TO} orientations. \cite{supplemental} Longitudinal and Hall resistivities were measured with a Quantum Design PPMS cryostat equipped with a 9 T superconducting magnet. First-principles calculations of band structures were performed with WIEN2k. \cite{supplemental, wien2k} We used the Perdew-Burke-Ernzerhof exchange-correlation functional \cite{PBE} and the augmented plane wave and local orbital (APW+lo) method including the spin-orbit coupling as implemented in the WIEN2k program. The Hall coefficient was derived within the semi-classical Boltzmann theory using the BoltzTraP code. \cite{supplemental, boltztrap}

\begin{figure}
\begin{center}
\includegraphics*[width=12.5cm]{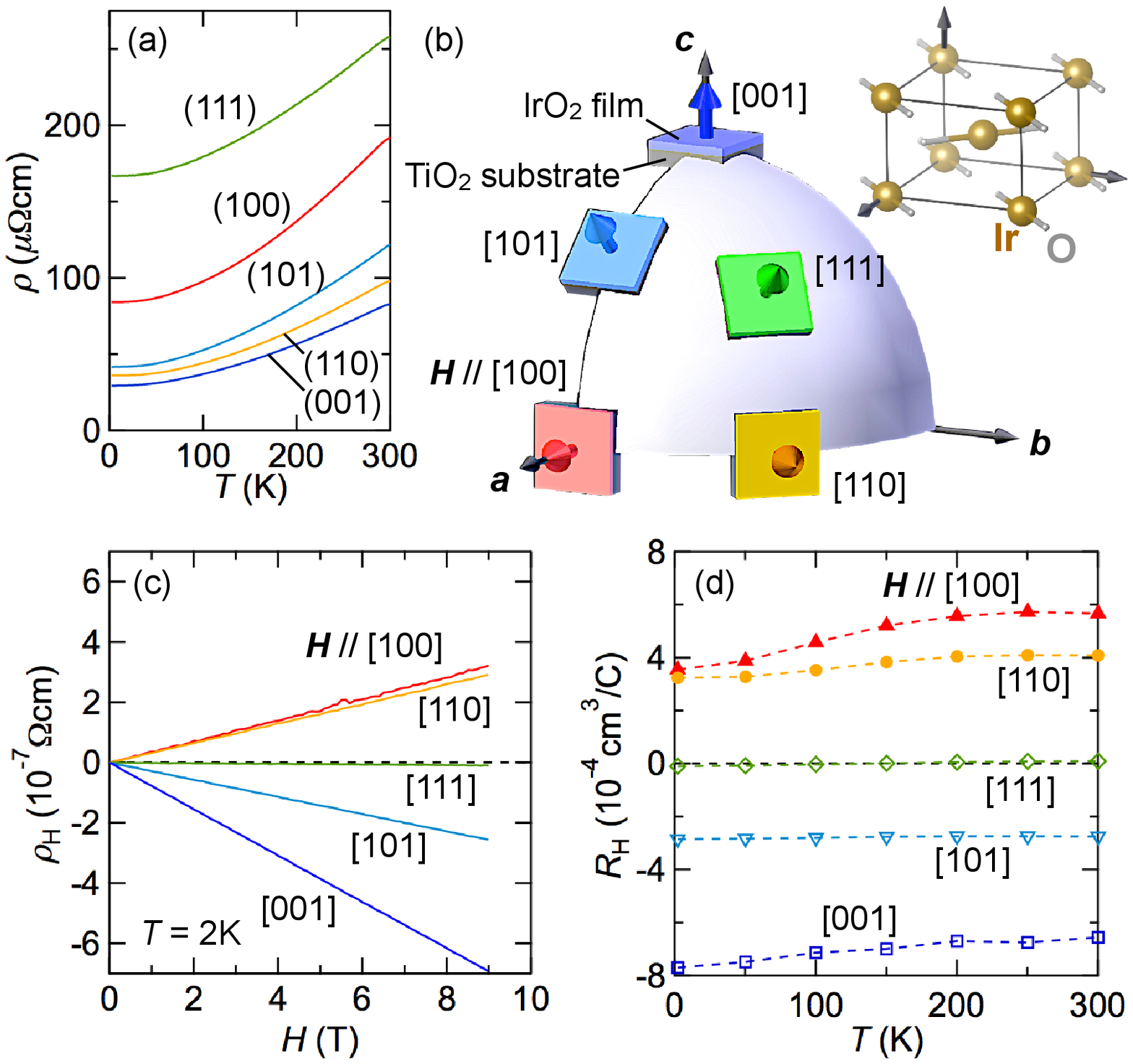}
\caption{
(Color online)
(a) Longitudinal resistivity of the series of the {\IO} oriented films. (b) Geometry of field directions and film orientations for the Hall measurement. {\IO} rutile structure is shown in the inset with the same orientation. (c) Field-direction dependence of the Hall resistivity measured at 2 K. (d) Temperature dependence of the ordinary Hall coefficient along the various field directions.
}
\label{fig3}
\end{center}
\end{figure}

In order to verify our prediction of the field-direction control of charge carriers, we have systematically measured the Hall effect for the series of the differently oriented {\IO} films. The film resistivity shown in Fig. 3(a) is dependent on the growth orientations and rather comparable to previously reported values of polycrystalline films, \cite{IrO2pld1, IrO2pld3, IrO2filmsputter1, IrO2ald1, IrO2ald2} probably stemming from the low-temperature growth process, with all samples indeed displaying metallic behavior down to the lowest temperature of 2 K. Geometry of magnetic field directions and film crystalline orientations for the Hall measurement is illustrated in Fig. 3(b). The ordinary Hall effect for $H \parallel [110]$, [100], [001], [101], and [111] is precisely quantified by measuring the films with the corresponding normal directions, in order to exclude complicated inclination angle dependence including the planar Hall effect. Figure 3(c) provides the Hall resistivity data taken for the five field directions at 2 K. The curves are almost entirely linear, with opposite slope depending on the field direction. While both types of the carriers have been partially observed with whisker crystals, \cite{IrO2transport1} our result clearly indicates that the dominant carrier type is switched uniquely with respect to the magnetic field direction; Holes mainly contribute to the electric conduction for $H \parallel [100]$ or [110], while electrons do for $H \parallel [001]$ or [101]. At zero field, transverse Hall voltage is not generated. But once a finite field is applied along a direction, a carrier type favorable for the fixed circular motion direction results in positive or negative Hall voltage. As shown in Fig. 3(d), each ordinary Hall coefficient $R_{\mathrm{H}}$, suitably derived from the linear field dependence, remains with the same sign up to room temperature, except for the borderline case of [111]. $R_{\mathrm{H}}$ for $H \parallel [111]$ is very close to zero due to the competition between the holes and electrons, showing a slight change from negative to positive with increasing temperature. 

\begin{figure}
\begin{center}
\includegraphics*[width=12.5cm]{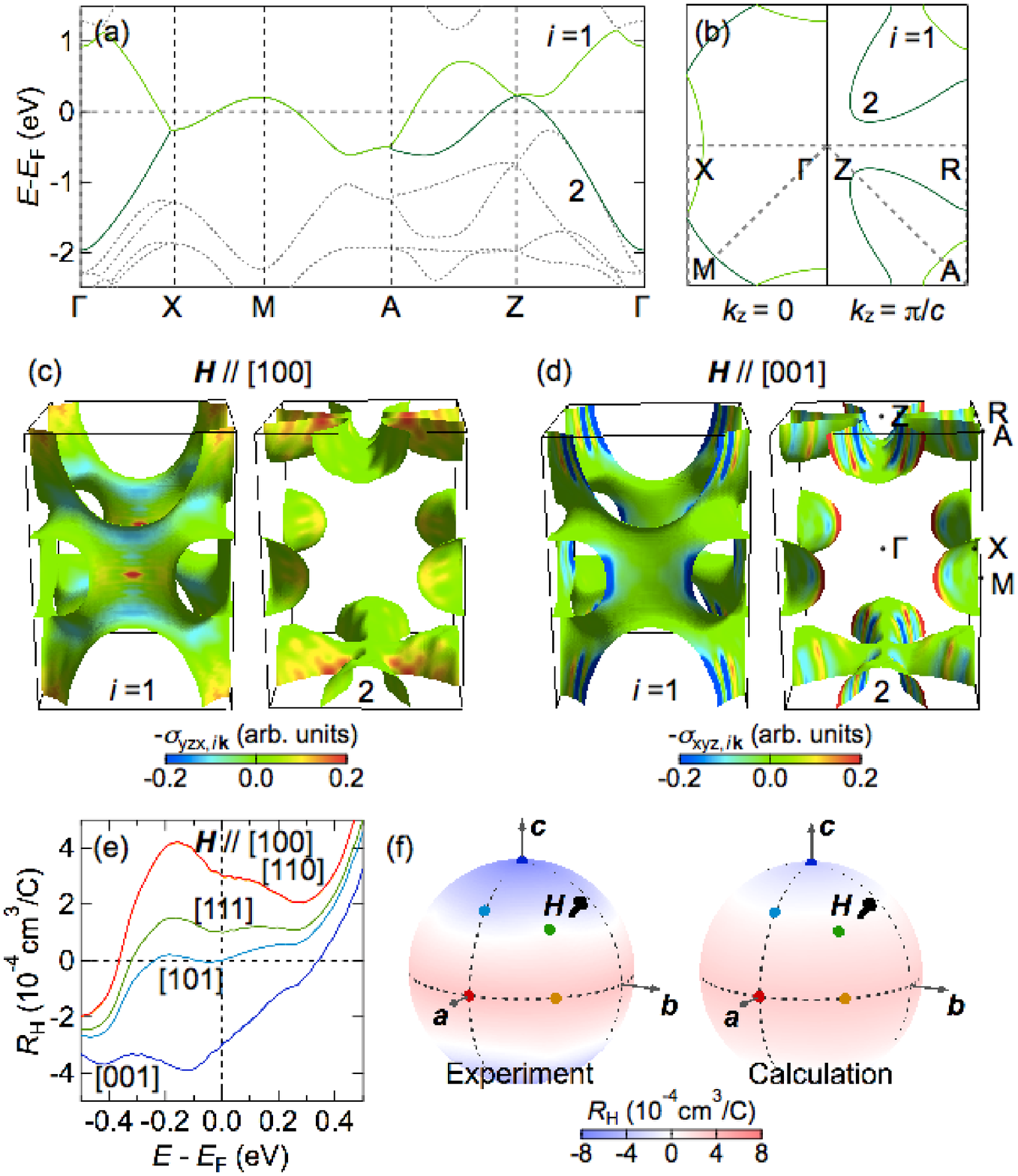}
\caption{
(Color online)
(a) Band structure and (b) in-plane Fermi surface cuts of {\IO}, calculated with bulk parameters. \cite{IrO2bulkstructure} Distribution of $\sigma_{\alpha\beta\gamma,i{\bf k}}$ on the two Fermi surfaces ($i =1, 2$), for the cases of applying the field along (c) [100] and (d) [001] directions. (e) Field-direction dependence of the ordinary Hall coefficient derived from the electronic structure. (f) Comparison between the experimental and theoretical values of the Hall coefficient, plotted for the field direction vector on the spherical coordinate. A dominant type of charge carriers is intrinsically switched from electron to hole with inclining the magnetic field from the $c$-axis onto the $ab$-plane. The five investigated directions for interpolation are represented by the colored dots. 
}
\label{fig4}
\end{center}
\end{figure}

Let us consider the influence of the low-energy band structure on the carrier transport in {\IO}. The calculated energy bands shown in Fig. 4(a) include two distinct bands 1 and 2 crossing the Fermi level, both of which are predominantly of the {\Jhalf} character of the Ir {\ttwog} states. \cite{IrO2calc3DMFT} Their fundamental features such as semi-metallic dispersions and bandwidth are largely similar to recent first-principles calculation results, \cite{IrO2bandopt, IrO2calc1DFT, IrO2calc2DFT, IrO2calc3DMFT} and are also in good agreement with photoemission and optical spectroscopy studies. \cite{IrO2bandopt, IrO2calc2DFT, IrO2XPS1, IrO2XPS2} In light of the nonsymmorphicity of {\IO}, the two bands stick together at high symmetry X($\pi, 0, 0$)--M($\pi, \pi, 0$)--A($\pi, \pi, \pi$) lines and Z ($0, 0, \pi$) points on the Brillouin zone boundary, originating from the [001] fourfold screw axes (M--A, Z) and the (100) glide planes (X--M, Z) in the rutile structure. \cite{nonsymmorphictext} The resultant folded Fermi surfaces are located near the boundary planes, as represented in Fig. 4(b). In particular, the in-plane cut at $k_{z}=0$ is nearly identical to the schematic Fermi surface picture expected in Fig. 1(c).

The anisotropy of the Hall coefficient can be visualized by plotting the momentum space distribution of the tensor $\sigma_{\alpha\beta\gamma,i{\bf k}}$ defined by the following relation.
\begin{eqnarray}
 \sigma_{\alpha\beta\gamma,i{\bf k}} = \epsilon_{\gamma\lambda\delta} v_{\alpha,i{\bf k}} v_{\lambda,i{\bf
  k}} M_{\beta\delta,i{\bf k}}^{-1}.
\end{eqnarray}
Here $\epsilon_{\gamma\lambda\delta}$ is the Levi-Civita symbol, and the group velocity $v_{\alpha,i{\bf k}}$ and the inverse mass tensor $M_{\beta\delta,i{\bf k}}^{-1}$ at the ${\bf k}$ point on the $i$-th band are calculated from smoothed Fourier interpolation of a mesh of band energies. \cite{supplemental, boltztrap} In Figs. 4(c) and 4(d), $\sigma_{\alpha\beta\gamma,i{\bf k}}$ on the two Fermi surfaces ($i =1, 2$) is plotted for the cases of applying the field along [100] and [001]. In stark contrast to a simple doped semiconductor or semimetal, $\sigma_{\alpha\beta\gamma,i{\bf k}}$ exhibits a nontrivial sign distribution over the Fermi surfaces and strong anisotropic nature for the field direction. For $H \parallel [100]$, contribution of holes (colored with red) is prominent near the boundary of $k_{z}=\pi /c$ on the surface 2. For $H \parallel [001]$, on the other hand, electrons (blue) seem more dominant especially near the $k_{x}=\pi /a$ boundary on the surface 1. The entire momentum space distribution of $\sigma_{\alpha\beta\gamma,i{\bf k}}$ is rather complicated, and symmetry relation between the field direction and tensor distribution may also depend on details of the band dispersions. The Hall coefficient for the various field directions is then uniquely derived from the electronic structure, as plotted against the band filling in Fig. 4(e). Its sign is critically dependent on the field direction around the Fermi level, and the trend agrees quite well with the experimental findings shown in Fig. 3(c). This conclusively proves that the observed field-direction dependent carrier type originates from the highly anisotropic electronic structure in nonsymmorphic {\IO}. Also, this finding should not be largely affected by band structure changes in strained films, \cite{supplemental} as the trend is suitably reproduced by the calculation using only bulk lattice parameters.

Figure 4(f) summarizes the measured and calculated Hall coefficient for the field direction on the spherical map. Overall, the dominant carrier type is mostly determined by the inclined angle $\theta_{H}$ between the field direction and the $c$-axis ($R_{\mathrm{H}} < 0$ for $0^{\circ} (H \parallel c) \leq \theta_{H} \lesssim 45^{\circ}$ and $R_{\mathrm{H}} > 0$ for $45^{\circ} \lesssim \theta_{H} \leq 90^{\circ}(H \parallel a)$). This sign relation of $R_{\mathrm{H}}$ to $\theta_{H}$ probably corresponds to the signs of predominant transfer integrals $t_{ab}, t_{ba}$, and $t_{c}$ in {\IO}, where $t_{ab}$ and $t_{ba}$ denote different diagonal transfers in the $ab$-plane. The sign of $R_{\mathrm{H}}$ will be entirely reversed for the transfer integrals of opposite sign, as confirmed in the tight binding model in Fig. 1(e). One minor difference between the experimental and theoretical results is that the charge carrier appears more hole-like in the calculation. This might be due to band- or momentum- dependent relaxation time, which is now assumed to be constant in the calculation. \cite{supplemental}

Here we present one direction of materials design for realizing the switching of the carrier type by external stimuli other than the electric field. In nonsymmorphic {\IO}, the Hall voltage can be reversed and also tuned even to zero depending on the crystal orientation, which may be useful for electronics and spintronics applications. Further studies will elucidate the detailed origins of the anisotropic responses on the Fermi surfaces and find potential technological applications of the unusual transport function.

In summary, in the quest to realize the field switching of the charge carrier type, we have built the simple tight-binding model and focused on the nonsymmorphic materials having the unique folded Fermi surfaces. We have prepared nonsymmorphic {\IO} single-crystalline films with various growth orientations, and systematically measured their Hall effect for the corresponding field directions. The results clearly demonstrate that the dominant carrier type is certainly controlled by the field direction. This switching intrinsically results from the electronic structure in {\IO}, as evidenced by the first-principles calculations showing the nontrivial $\sigma_{\alpha\beta\gamma,i{\bf k}}$ distribution on the Fermi surfaces and its strong field direction dependent anisotropic nature. Our study will inspire further investigations of the switching function and other intriguing electronic features in the nonsymmorphic conductive materials. 

This work was partly supported by JSPS Grants-in-Aid for Scientific Research(S) (No. 24226002) and Challenging Exploratory Research (No. 26610098). We thank J. Falson for proofreading of the manuscript.


\begin{thebibliography}{100}
\bibitem{mott} M. Imada, A. Fujimori, and Y. Tokura, Rev. Mod. Phys. \textbf{70,} 1039 (1998).
\bibitem{ti1} M. Z. Hasan and C. L. Kane, Rev. Mod. Phys. \textbf{82,} 3045 (2010).
\bibitem{ti2} X.-L. Qi and S.-C. Zhang, Rev. Mod. Phys. 83, 1057 (2011).
\bibitem{nonsymmorphic1} S. A. Parameswaran, A. M. Turner, D. P. Arovas, and A. Vishwanath, Nat. Phys. \textbf{9,} 299 (2013).
\bibitem{nonsymmorphic2} C.-X. Liu, R.-X. Zhang, and B. K. VanLeeuwen, Phys. Rev. B \textbf{90,} 085304 (2014).
\bibitem{nonsymmorphictext} V. Heine, \textit{Group Theory of Quantum Mechanics} (Pergamon, New York, 1960).
\bibitem{ambipolar} M. Muccini, Nat. Mater. \textbf{5,} 605 (2006).
\bibitem{IrO2transport1} W. D. Ryden, A. W. Lawson, and C. C. Sartain, Phys. Lett. \textbf{26A,} 209 (1968).
\bibitem{SOI} B. J. Kim, H. Ohsumi, T. Komesu, S. Sakai, T. Morita, H. Takagi, and T. Arima, Science \textbf{323,} 1329 (2009).
\bibitem{IrO2bandopt} J. S. de Almeida and R. Ahuja, Phys. Rev. B \textbf{73,} 165102 (2006).
\bibitem{IrO2XPS1} G. K. Wertheim and H. J. Guggenheim, Phys. Rev. B \textbf{22,} 4680 (1980).
\bibitem{IrO2XPS2} R. R. Daniels, G. Margaritondo, C.-A. Georg, and F. Levy, Phys. Rev. B \textbf{29,} 1813 (1984).
\bibitem{IrO2XAS} J. P. Clancy, N. Chen, C. Y. Kim, W. F. Chen, K. W. Plumb, B. C. Jeon, T. W. Noh, and Y.-J. Kim, Phys. Rev. B \textbf{86,} 195131 (2012).
\bibitem{IrO2RXS} Y. Hirata, K. Ohgushi, J. Yamaura, H. Ohsumi, S. Takeshita, M. Takata, and T. Arima, Phys. Rev. B \textbf{87,} 161111(R) (2013).
\bibitem{IrO2calc2DFT} J. M. Kahk, C. G. Poll, F. E. Oropeza, J. M. Ablett, D. C\'{e}olin, J-P. Rueff, S. Agrestini, Y. Utsumi, K. D. Tsuei, Y. F. Liao, F. Borgatti, G. Panaccione, A. Regoutz, R. G. Egdell, B. J. Morgan, D. O. Scanlon, and D. J. Payne, Phys. Rev. Lett. \textbf{112,} 117601 (2014).
\bibitem{IrO2calc1DFT} M.-S. Miao and R. Seshadri, J. Phys. Condens. Matter \textbf{24,} 215503 (2012).
\bibitem{IrO2calc3DMFT} S. K. Panda, S. Bhowal, A. Delin, O. Eriksson, and I. Dasgupta, Phys. Rev. B \textbf{89,} 155102 (2014).
\bibitem{rutileoxides1} J. B. Goodenough, J. Solid State Chem. \textbf{3,} 490 (1971).
\bibitem{rutileoxides2} L. F. Mattheiss, Phys. Rev. B \textbf{13,} 2433 (1976).
\bibitem{rutileoxides3} J. H. Xu, T. Jarlborg, and A. J. Freeman, Phys. Rev. B \textbf{40,} 7939 (1989).
\bibitem{IrO2pld1} M. A. El Khakani, M. Chaker, and E. Gat, Appl. Phys. Lett. \textbf{69,} 2027 (1996).
\bibitem{IrO2pld3} C. Wang, Y. Gong, Q. Shen, and L. Zhang, Appl. Surf. Sci. \textbf{253,} 2911 (2006).
\bibitem{IrO2filmsputter1} T. Ishikawa, Y. Abe, M. Kawamura, and K. Sasaki, Jpn. J. Appl. Phys. \textbf{42,} 213 (2003).
\bibitem{IrO2ald1} S. W. Kim, S. H. Kwon, D. K. Kwak, and S.-W. Kang, J. Appl. Phys. \textbf{103,} 023517 (2008).
\bibitem{IrO2ald2} J. H\"{a}m\"{a}l\"{a}inen, M. Kemell, F. Munnik, U. Kreissig, M. Ritala, and M. Leskel\"{a}, Chem. Mater. \textbf{20,} 2903 (2008).
\bibitem{IrO2pld2} B. R. Chalamala, Y. Wei, R. H. Reuss, S. Aggarwal, B. E. Gnade, R. Ramesh, J. M. Bernhard, E. D. Sosa, and D. E. Golden, Appl. Phys. Lett. \textbf{74,} 1394 (1999).
\bibitem{IrO2filmsputter2} H. W. Kim, S. H. Shim, J. H. Myung, and C. Lee, Vacuum, \textbf{82,} 1400 (2008).
\bibitem{IrO2SH} K. Fujiwara, Y. Fukuma,	 J. Matsuno, H. Idzuchi, Y. Niimi, Y. Otani, and H. Takagi, Nat. Comm. \textbf{4,} 2893 (2013).
\bibitem{supplemental} See Supplemental Material at [URL will be inserted by publisher] for more information regarding film characterization and calculations.
\bibitem{wien2k} P. Blaha, K. Schwarz, G. K. H. Madsen, D. Kvasnicka, and J. Luitz, \textit{WIEN2K, An Augmented Plane Wave + Local Orbitals Program for Calculating Crystal Properties} (Techn. Universit\"{a}t Wien, Austria, 2001).
\bibitem{PBE} J. P. Perdew, K. Burke, and M. Ernzerhof, Phys. Rev. Lett. \textbf{77,} 3865 (1996).
\bibitem{boltztrap} K. H. Madsen and D. J. Singh, Comput. Phys. Commun. \textbf{175,} 67 (2006).
\bibitem{TiO2bulkstructure} C. J. Howard, T. M. Sabine, and F. Dickson, Acta Crystallogr. B \textbf{47}, 462 (1991).
\bibitem{IrO2bulkstructure} A. A. Bolzan, C. Fong, B. J. Kennedy, and C. J. Howard, Acta Crystallogr. B \textbf{53}, 373 (1997).
\newpage
\end{thebibliography}
\end{document}